\documentclass[12pt]{iopart}

\usepackage{iopams}  
\usepackage{graphicx}
\expandafter\let\csname equation*\endcsname\relax
\expandafter\let\csname endequation*\endcsname\relax
\usepackage{amsmath}
\usepackage{color} 
\usepackage{hyperref}
\hypersetup{
colorlinks=true,final=true,
        linkcolor=blue,
        citecolor=blue,
        filecolor=blue,
        urlcolor=blue,
}

\begin{document}

\title{Comparative study of the compensated semi-metals LaBi and LuBi : A first-principles approach}
\author{Urmimala Dey}
\address{Centre for Theoretical Studies, Indian Institute of Technology Kharagpur, Kharagpur-721302, India}
\ead{urmimaladey@iitkgp.ac.in}
\vspace{10pt}

\begin{abstract}
 We have investigated the electronic structures of LaBi and LuBi, employing the full-potential all
electron method as implemented in Wien2k.
Using this, we have studied in detail both the bulk and the surface states of these materials.
From our band structure calculations we find that LuBi, like LaBi, is a compensated semi-metal with
almost equal and sizable electron and hole pockets.  In analogy with experimental evidence
in LaBi, we thus predict that LuBi will also be a candidate for extremely large magneto-resistance (XMR),
which should be of immense technological interest. Our calculations reveal that LaBi, despite being
gapless in the bulk spectrum, displays the characteristic features of a $\mathbb{Z}_{2}$
topological semi-metal, resulting in gapless Dirac cones on the surface, whereas LuBi only shows avoided band inversion in the bulk and is thus a
conventional compensated semi-metal with extremely large magneto-resistance.
\end{abstract}

%
%
%
%
%

\section{Introduction}
The discovery of quantum Hall effect~\cite{LuBi1} provided the first example of a quantum state, whose explanation
necessitates a scheme beyond Landau's approach of spontaneous symmetry breaking and hence leads to topological classification.
Topological insulators (TI)~\cite{LuBi2,LuBi3} are characterized by a bulk electronic band structure that resembles an ordinary
band insulator, but the surface contains conducting states which are protected against perturbations as long as time-reversal and/or other spatial symmetries are unbroken. Topological states can also be realized in Dirac and Weyl semi-metals~\cite{LuBi4,LuBi5,LuBi6,LuBi7,LuBi8} as a result of linear
band crossings between conduction and valence bands in the bulk spectrum. The surface states of these topological semi-metals (TSM) are characterized by
Fermi arcs~\cite{LuBi4} and they generally possess high-carrier mobilities and high
magneto-resistance (MR)~\cite{LuBi9,LuBi10}. The stability of these topological phases requires the presence of certain symmetries.
A Dirac point, for instance, is stable only in the presence of time-reversal and space inversion symmetry, and splits into a pair of Weyl points with opposite chiralities if either of these symmetries is broken. 

There also exits a class of gapless topological systems which have recently been dubbed
$\mathbb{Z}_{2}$ topological semi-metals~\cite{LuBi11}. Although gapless in the bulk, these systems are characterized by a non-trivial $\mathbb{Z}_{2}$ invariant, which require the presence of TRS for the stabilization of their non-trivial topological properties. It is important to recognize that in these systems the bulk spectrum is gapless, with electron and hole pockets as in a compensated semi-metal, but the existence of a direct gap at each value of $k$ allows the definition of a $\mathbb{Z}_{2}$ invariant. However, even with a non-trivial $\mathbb{Z}_{2}$ invariant, because of a gapless bulk spectrum, in these systems the existence of odd number of Dirac cones on the surfaces can only be accidental.

Recently, the family of binary rare-earth monopnictides REX (RE = La, Y, Nd, Ce and X = Sb, Bi)
have stimulated interest, due to their novel topological states along with extremely large magneto-resistance (XMR)~\cite{LuBi12,LuBi13,LuBi14,LuBi15,LuBi16,LuBi17,LuBi18}.
 Among them, LaBi and LaSb have been classified as $\mathbb{Z}_{2}$ topological semi-metals based on both theoretical and
experimental findings~\cite{LuBi11,LuBi19,LuBi20,LuBi21,LuBi22}. Moreover, the discovery of non-saturating quadratic magneto-resistance
of extremely large magnitude in LaBi and LaSb~\cite{LuBi22,LuBi23,LuBi24,LuBi25} has made them promising candidates for device
applications.

The electronic, structural, mechanical and thermal properties of lutetium monopnictides (LuX : X = N, P, As, Sb, Bi) have been
studied by different groups since 2010~\cite{LuBi26,LuBi27,LuBi28,LuBi29,LuBi41}. However, not much significant experimental or theoretical works
 exist on the transport and topological properties of Lu-monopnictides. Recently, Pavlosiuk \textit{et al}~\cite{LuBi42} have identified LuSb as a high magnetoresistive material due to the compensation of the electron and hole contributions. The last member of this family LuBi is of particular
 interest because the effect of the spin-orbit coupling (SOC) is strongest for LuBi, which may result in non-trivial topological
 properties.

In this work, we have performed a comparative study of the electronic structures of LaBi and LuBi using all-electron
density functional theory (DFT) calculations. The valence electronic configurations of both the materials
are same except for the fact that the 4f orbital of Lu is completely filled in LuBi whereas LaBi contains an
empty La-4f orbital. The bulk band structures calculated using Local Density Approximation (LDA) show a band inversion
near the $X$-point with electron and hole pockets at the Fermi level in both cases. We then use the modified Becke Johnson (mBJ) potential to further verify the stability of the band inversion near the $X$-point, which holds the key to the topological nature
of LaBi and LuBi. We find that there is no
qualitative change in the band structure of LaBi when the mBJ functional is used; however, the band inversion is
further consolidated on the introduction of mBJ. On the contrary, in case of LuBi, the band inversion disappears
and it is converted to an avoided band inversion between the $\Gamma$ and $X$ points in the Brillouin zone. However, we can still detect direct energy gap at each $k$-point which allows us to calculate
the $\mathbb{Z}_{2}$ invariant~\cite{LuBi11} for LaBi and LuBi. Calculation of the first $\mathbb{Z}_{2}$ index $\nu_0$ shows that LaBi possesses a non-zero $\mathbb{Z}_{2}$ topological invariant, whereas in LuBi $\nu_0$ is zero, indicating the topologically trivial nature of LuBi. Moreover, we identify one gapped Dirac cone at the $ \bar{\Gamma} $ point
and one gapless Dirac cone at the $\bar{M}$ point in the (001) surface band structure of LaBi~\cite{LuBi20,LuBi21}. However, the (001) and (111) surface band structures of LuBi show that no Dirac cone exists on the (001) and (111) surfaces.
More interestingly, the derived ratio of the number of electron and hole carriers in LaBi and LuBi
leads us to conclude that they are electron-hole compensated semi-metals, which could give rise to extremely large
magneto-resistance (XMR) as expected from the semiclassical two-band model~\cite{LuBi24,LuBi25}. Our calculations thus
reveal that LaBi, despite being gapless in the bulk spectrum, displays the characteristic features of a $\mathbb{Z}_{2}$
topological semi-metal, whereas LuBi shows avoided band inversion in the bulk and is thus a conventional compensated
semi-metal with the possibility of extremely large magneto-resistance, which is of immense technological importance.

\section{Methodology}
Our $\textit{ab initio}$ calculations are performed using the full-potential linearized augmented plane wave (FLAPW)
approach as implemented in the WIEN2K code~\cite{LuBi30}. The local density approximation (LDA) and the modified Becke Johnson (mBJ) potential~\cite{LuBi31,LuBi32} are employed for the exchange-correlation. Effect of Spin-orbit coupling (SOC) is also included. We find that there is no qualitative change
in the band structure when a small Hubbard correlation term (U = 0.25 eV) is included in the calculations. For this reason, the calculations
presented here are carried out within the LDA+SO scheme. A $ 21 \times 21 \times 21 $ k-point mesh is used in the full Brillouin zone for the scf
calculation. We have taken $ L_{max} = 12 $ for the partial waves inside the spheres and $ G_{max} = 14 $ for the charge Fourier
expansion. For the interstitial regions, the plane wave cut-off is set to $ K_{max} $ = $ \frac{9.5}{RMT} $, where $ K_{max} $ is
the largest k-vector used in the plane-wave expansion and RMT denotes the smallest atomic sphere radius. The cell parameters are
optimized to get the equilibrium lattice constant. We use the LDA functional to account for the exchange correlation,
because, for heavier elements LDA is believed to yield more accurate values of equilibrium lattice parameter ($a_0$) and
bulk modulus ($B$)~\cite{LuBi32a}. To that end, we  calculate $a_0$ and $B$ for LaBi using three different
exchange correlation functionals, namely GGA-PBE, GGA-WC and LDA. The calculations with LDA functional achieve both the
quantities nearest to the reported experimental values~\cite{LuBi32b}. We relax the atomic co-ordinates for both bulk and surface band structure calculations until each force component on the atoms is less than 1 mRy/Bohr. 

In order to get the Maximally Localized Wannier Functions (MLWFs) of La/Lu d-$ t_{2g} $ and Bi p orbitals, we use the WANNIER90
package~\cite{LuBi33,LuBi34,LuBi35}, which is then used to calculate the Fermi-surface and the carrier concentrations. A
$ 40 \times 40 \times 40 $ k-mesh is used for the wannier calculation.

\section{Results and Discussion}
\subsection{Bulk band structure}
\begin{figure}[h]
\centering
$\begin{array}{cc}
\includegraphics[scale=.22]{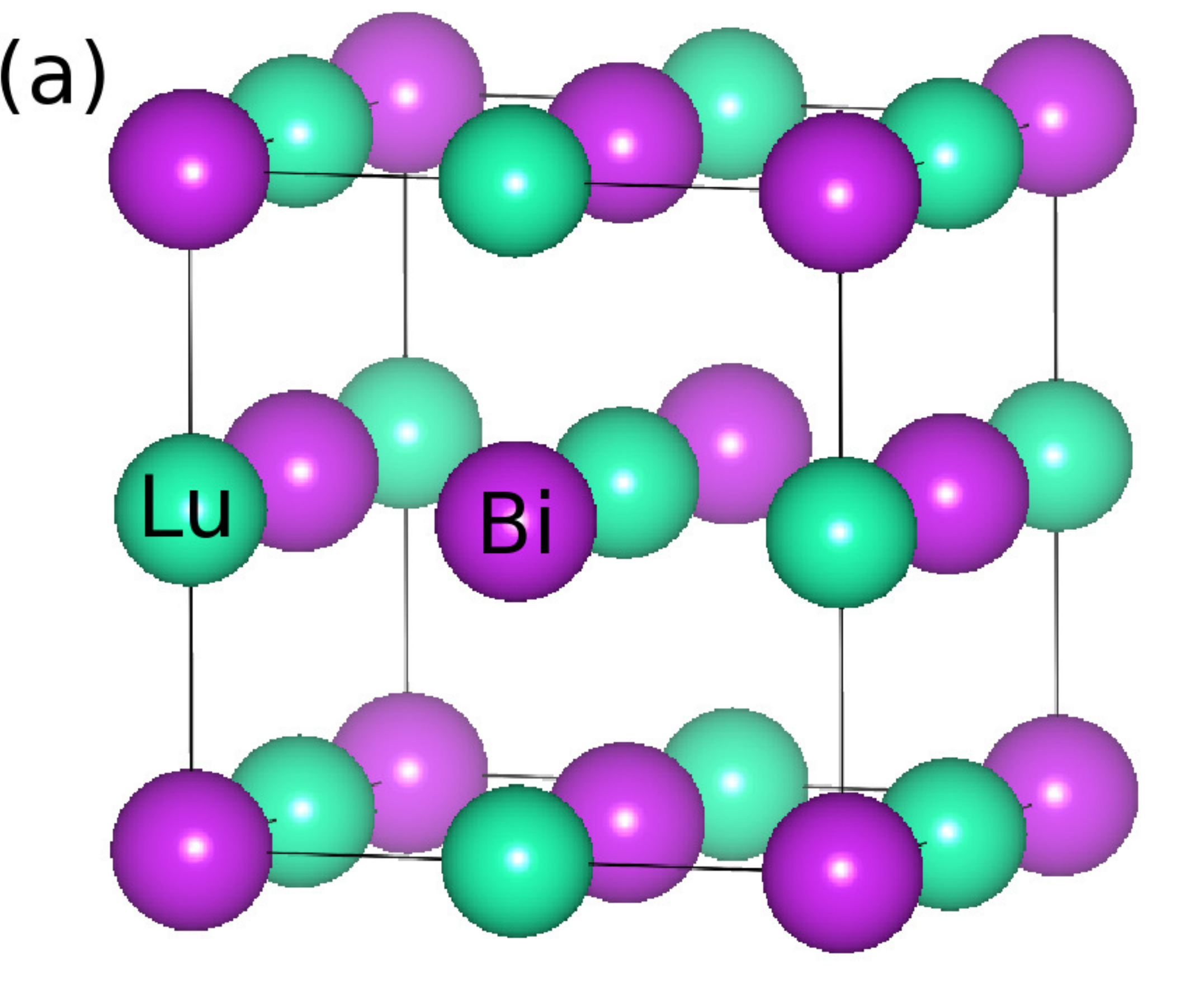} &
\includegraphics[scale=.25]{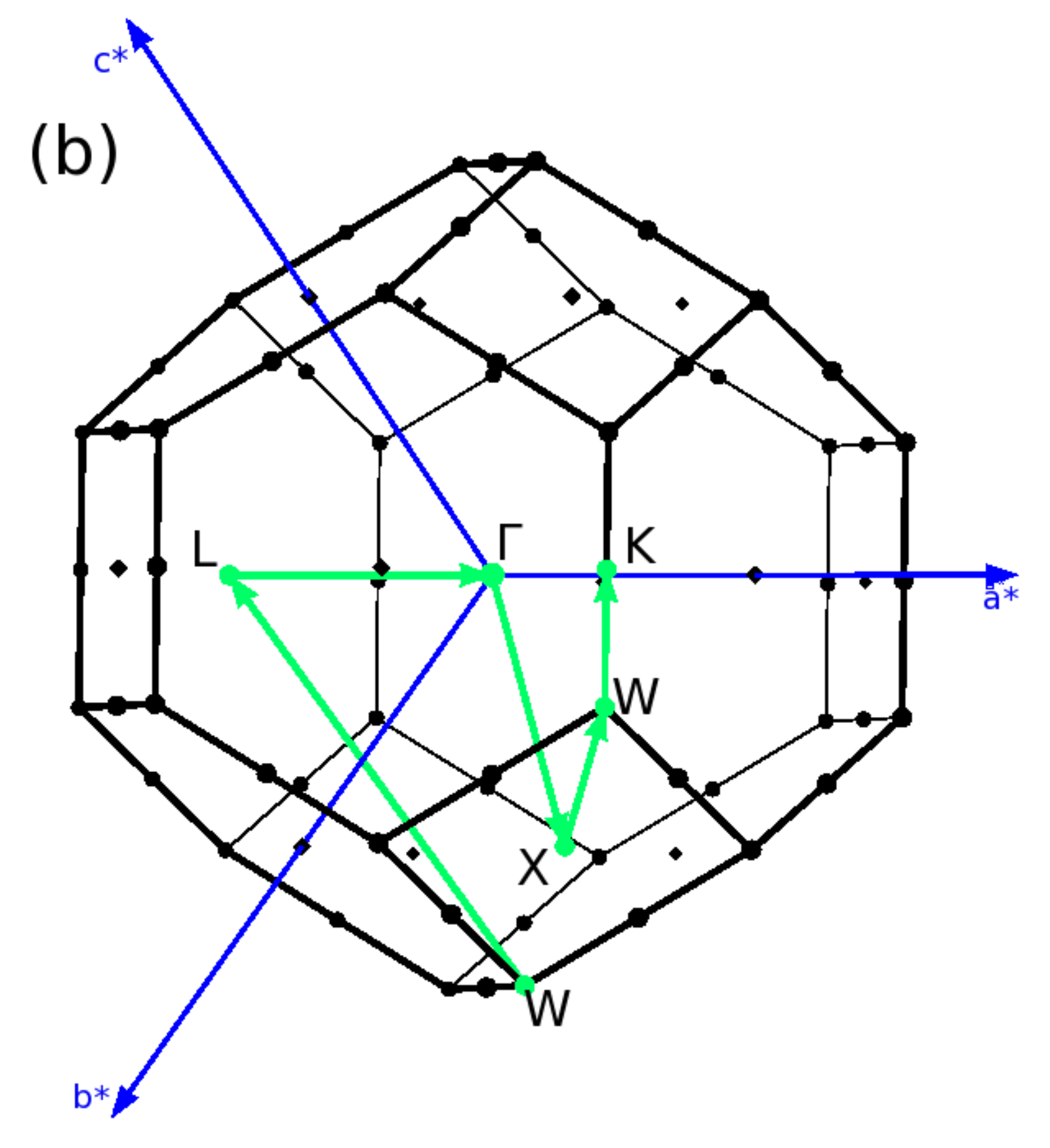} \\
\end{array}$
\caption{(a) The face-centered cubic structure of LuBi. The cyan spheres indicate the Lu atoms and the Bi atoms are indicated by the purple spheres. (b) The bulk Brillouin zone of the fcc lattice with high symmetry points. $ a^{*}, b^{*}, c^{*} $ are reciprocal lattice vectors.}
\label{struct}
\end{figure}

Both LaBi and LuBi crystallize into a face-centered cubic structure with space group symmetry $Fm\bar{3}m$ , in which the Bi atom is positioned
at ($0$,$0$,$0$) and La/Lu is located at ($ \frac{1}{2}, \frac{1}{2}, \frac{1}{2} $), as shown in Fig.~\ref{struct}(a). Fig.~\ref{struct}(b)
displays the bulk BZ of the fcc lattice. 

\begin{figure}[h]
\centering
\includegraphics[scale=0.28]{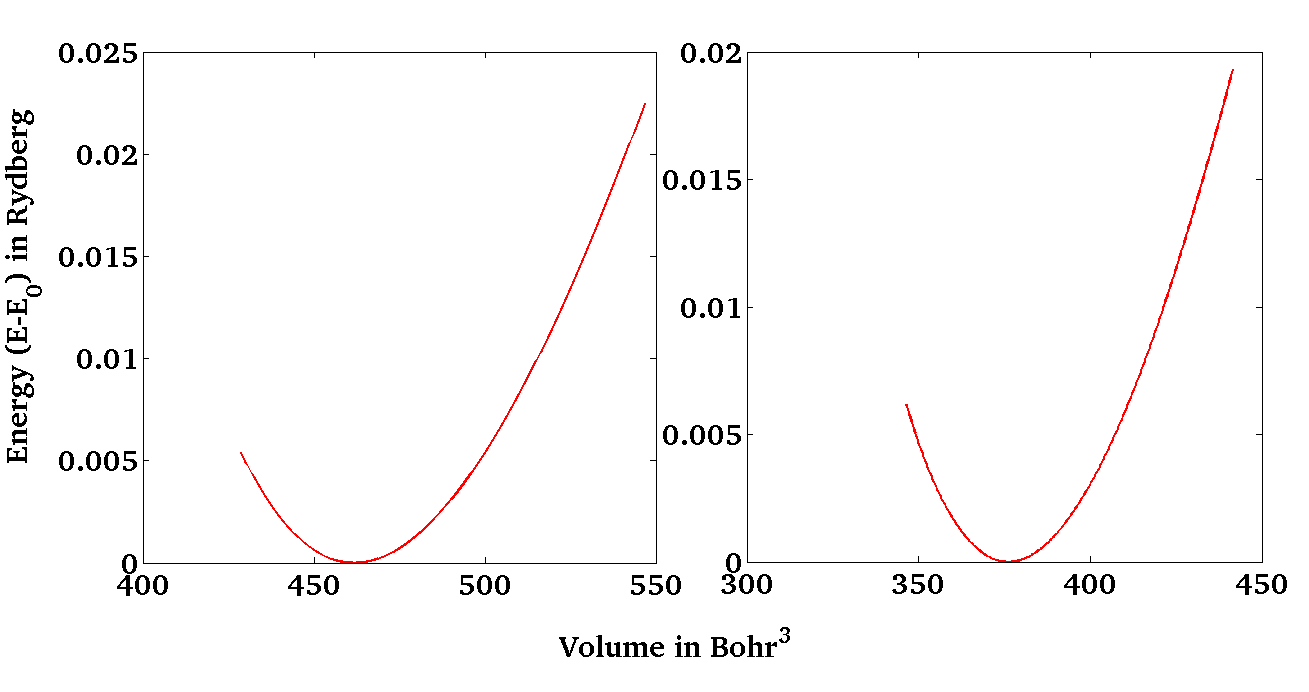}
\caption{Energy vs volume curves obtained from LDA calculation. LaBi (left) and LuBi (right)}
\label{optimization}
\end{figure}

Optimizing the
total energy of the system as a function volume and fitting the data to the Birch-Murnaghan (BM) equation~\cite{LuBi36,LuBi37},
we obtain the equilibrium lattice parameters and the bulk moduli. Fig.~\ref{optimization} shows the calculated and BM fitted energies
for LaBi (left) and LuBi (right).
The calculated lattice constants $a_{0}$, Bulk moduli (B) and its pressure derivatives $ B^{\prime} $ ( = $\frac{\partial B}{\partial P}$ ) are summarized in Table \ref{tab1}.

\begin{table}[h!]
\centering
\caption{The equilibrium lattice constants $a_{0}$, Bulk moduli (B) and its pressure derivatives $ B^{\prime} $ obtained by fitting data to the Birch-Murnaghan equation}
\label{tab1}
\begin{tabular}{c c c c}
\hline
   & $a_{0}$ (\r{A}) &  $B$ (GPa) & $ B^{\prime} $ \\
    \hline
    LaBi & 6.4927 & 58.4351 & 4.7534 \\
    LuBi & 6.0639 & 66.2371 & 4.2821 \\
    \hline
\end{tabular}
\end{table}

\subsubsection{bulk LaBi}
\begin{figure}[h]
\centering
 $\begin{array}{ccc}
\includegraphics[scale=.32]{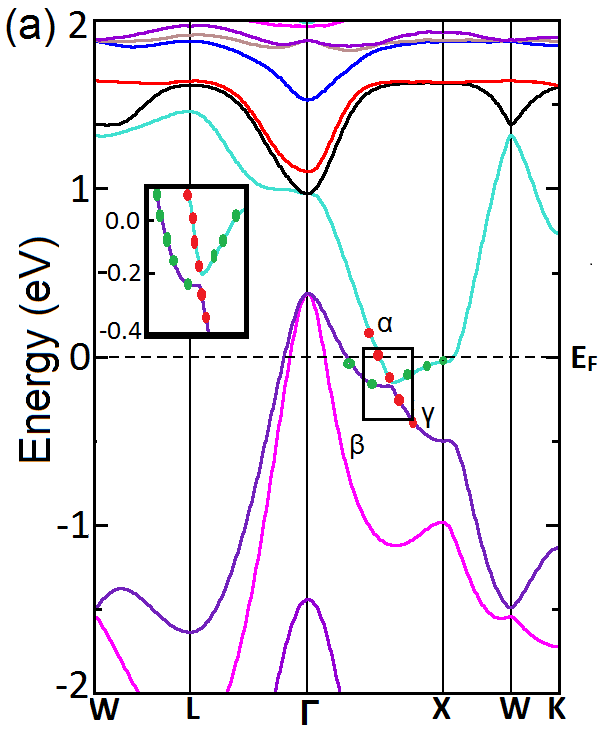}  &
\includegraphics[scale=.32]{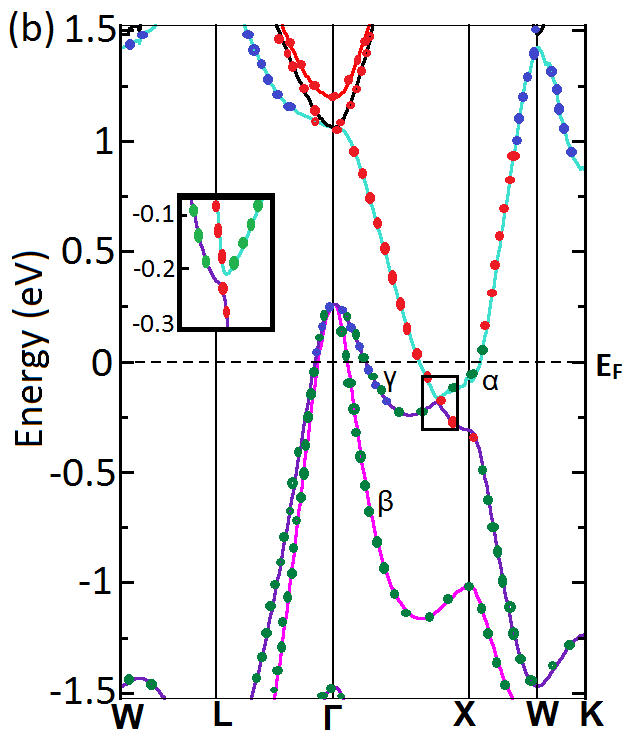} &
\includegraphics[scale=.32]{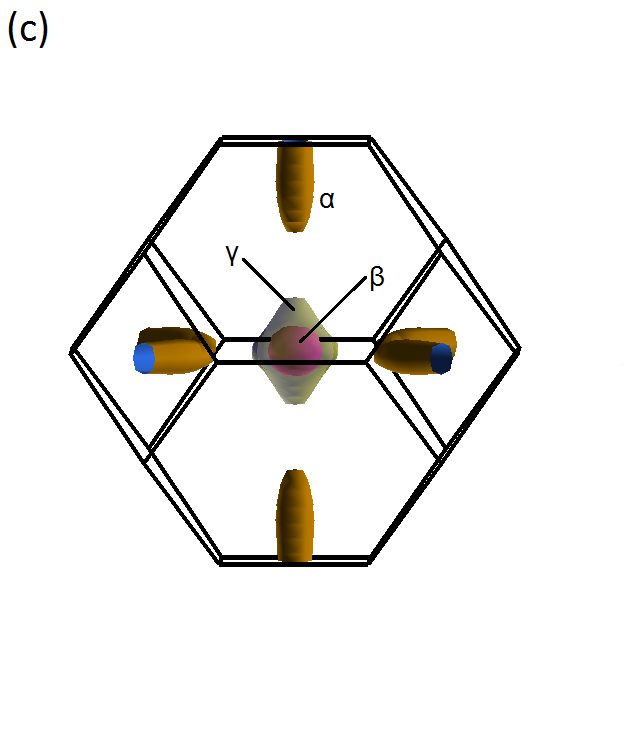} \\
\end{array}$
\caption{(a) Calculated band structures of LaBi with LDA functional along the $k$-path shown in \ref{struct}(b) : (a) using LDA functional and (b) using mBJLDA functional. When the SOC is taken into account a band inversion appears near the X-point, which is further consolidated with the introduction of the mBJ potential. Here, the red dots indicate the contribution of La d orbitals, the blue dots show the La-f contribution and the green dots show the contribution of the Bi-p orbitals. (c) The 3D Fermi surface of LaBi showing the electron ($\alpha$) and hole pockets ($\beta$ and $\gamma$).}
\label{bulk_LaBi}
\end{figure}

The band structure calculation of bulk LaBi along the high symmetry directions shown in Fig.~\ref{struct}(b) using LDA functional and including SOC shows that three doubly degenerate bands ($\alpha, \beta, \gamma$) cross the Fermi level $E_F$, creating two hole pockets $(\beta $ and $ \gamma )$ at the $\Gamma$-point and one electron pocket $(\alpha)$ near the X-point. This indicates the ambipolar nature of LaBi. The bulk band structure of LaBi calculated using the LDA functional is shown in Fig.~\ref{bulk_LaBi}(a). The calculated 3D Fermi surface is shown in \ref{bulk_LaBi}(c). As can be seen, electron pockets are centered around the X points and there
are two hole pockets around the $\Gamma$ point. Moreover, it is found that the conduction and valence band get inverted along the $ \Gamma-X $ direction i.e. the orbital character of the bands changes after crossing the point of inversion near the $X$-point~\cite{LuBi11,LuBi19,LuBi20,LuBi21,LuBi22}. Contribution to this band inversion comes from the $d_{xy}$ orbital of La and $p_{x}$ and $p_{y}$ orbitals of Bi. At each k-point the conduction and valence bands are gapped, although there is no indirect gap present in the band structure. The energy gap between the conduction and valence bands $\sim$ 19 meV at the point of inversion, shown in the inset of Fig.~\ref{bulk_LaBi}(a).

To confirm the topological character of LaBi, we revisit the bulk band structure using mBJ functional and re-evaluate the band inversion in the $\Gamma$-$X$ direction. The mBJ meta-GGA functional has proved its utility in correcting the band gap of insulators and semiconductors~\cite{LuBi37a}. However, its application to
semi-metals is  relatively new and has recently been applied on systems of topological importance~\cite{LuBi25,LuBi37b}. The LDA+mBJ+SO bulk
electronic structure calculations (Fig.~\ref{bulk_LaBi}(b)) reveal that the band inversion is not only preserved in LaBi but also further consolidated by increasing the band overlap, in agreement with previous first-principles calculations of LaBi~\cite{LuBi25,LuBi43}. The gap between the valence and the conduction band at the point of inversion is significantly reduced ( $\sim$ 15 meV) on the application of mBJ. However, the conduction and valence band remain gapped at each $k$-point. This allows one to determine the $\mathbb{Z}_{2}$ invariant for LaBi.

Unlike the two dimensional Quantum Spin Hall insulators, there are four $\mathbb{Z}_2$ invariants in three dimensions which distinguish the two kinds of topological classes : strong (STI) and weak topological insulators (WTI). The WTIs are not stable and can be destroyed by disorder. On the other hand, the STIs are stable against time-reversal-preserving disorders and give rise to novel topological conducting surface states.

For a material possessing both time-reversal and inversion symmetry, the $\mathbb{Z}_2$ topological invariants can be calculated by taking the parity product of the filled bands at the time-reversal-invariant momentum (TRIM) points using the Fu and Kane formula~\cite{LuBi45}. Here, we determine the first $\mathbb{Z}_2$ index $\nu_0$ which indicates whether the material is a STI or WTI. In 3D, there are eight TRIM points and $\nu_0$ is defined as
\begin{equation}
{(-1)}^{\nu_0} = \prod_{m=1}^{8} \delta_m
\label{Z2}
\end{equation}

where, $\delta_m$ is the parity product of the filled bands at the $m$-th TRIM point.
\begin{table}[h]
  \centering
  \caption{Calculation of $\mathbb{Z}_2$ index for LaBi and LuBi using Eq.~\ref{Z2}. As can been seen, LaBi possesses a non-trivial $\mathbb{Z}_2$ index $\nu_0$ = 1, whereas, it is zero for LuBi, indicating the topologically trivial nature of LuBi.}
  \label{tab2}
  \begin{tabular}{|c c |c|c c |c|}
    \hline
    \multicolumn{3}{|c|}{LaBi}& \multicolumn{3}{c|}{LuBi} \\
    \hline
     TRIM points & $\delta_m$ & $\nu_0$ & TRIM points & $\delta_m$ & $\nu_0$\\
    \hline
    1$\Gamma$ & $+$ & &1$\Gamma$ & $+$ & \\
    4$L$  & $+$ & 1 & 4$L$ & $+$ & 0\\
    3$X$ & $-$ & & 3$X$ & $+$ &\\
    \hline
\end{tabular}
\end{table}

The bands other than the $\alpha$, $\beta$ and $\gamma$ bands are isolated and topologically trivial. So, only the $\alpha$, $\beta$ and $\gamma$ bands are considered to determine the $\mathbb{Z}_2$ index. We calculate the first $\mathbb{Z}_2$ invariant $\nu_0$ and find that $\nu_0$ is 1 for LaBi which is in accordance with the previous calculations~\cite{LuBi11} and indicates the topologically non-trivial nature of LaBi. Calculation of $\nu_0$ is given in Table \ref{tab2}.
\subsubsection{Bulk LuBi}
\begin{figure}[h]
\centering
 $\begin{array}{ccc}
\includegraphics[scale=.32]{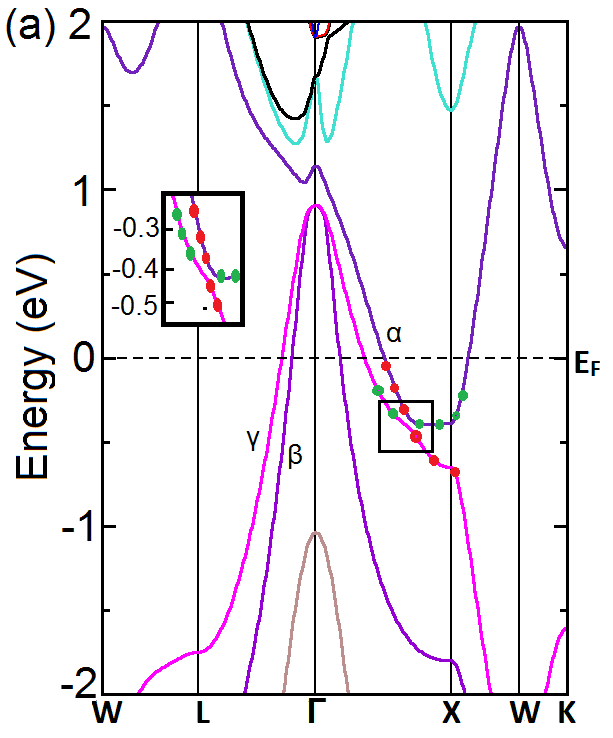} &
\includegraphics[scale=.32]{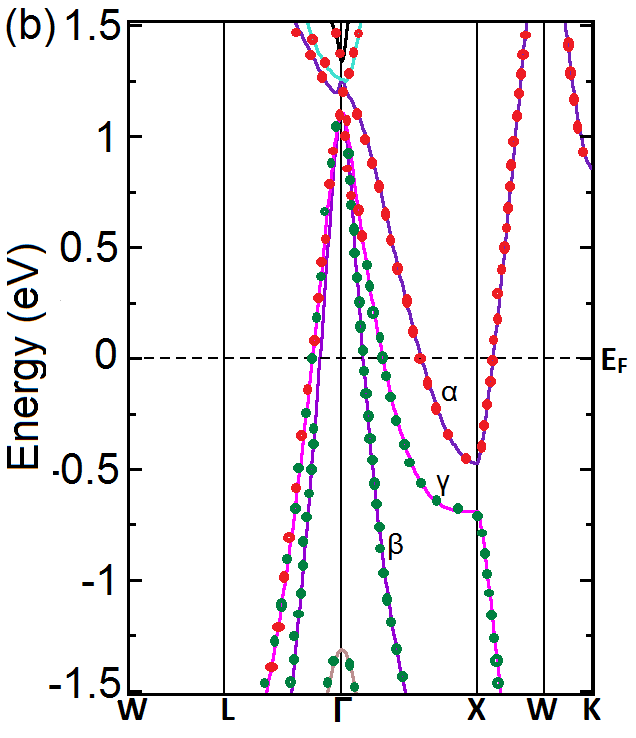} &
\includegraphics[scale=.32]{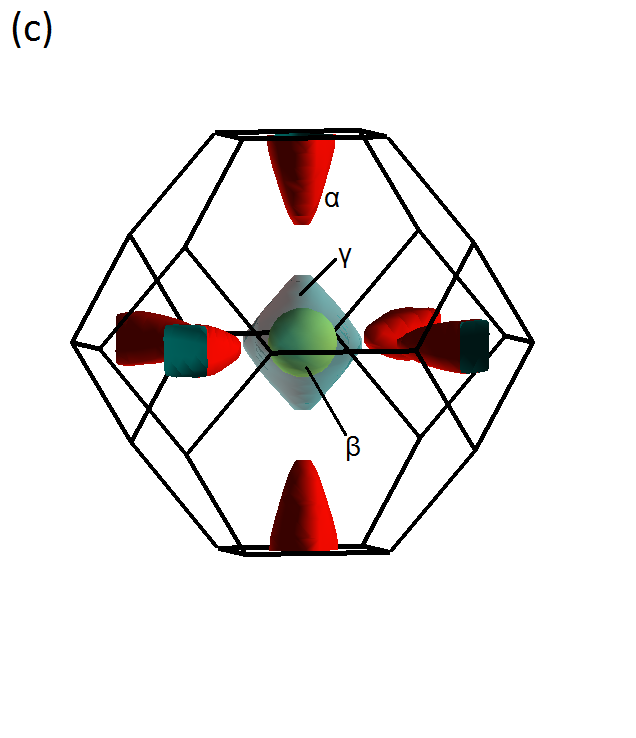} \\
\end{array}$
\caption{The bulk band structures LuBi with orbital weights, calculated including SOC and using (a) LDA functional (b) mBJLDA functional. The red dots indicate the contribution of the d orbitals of Lu, the green dots give the contribution of Bi-p orbitals. The band inversion which was present in the LDA calculation is obliterated by the introduction of the mBJ potential. (c) 3D Fermi surface plot of LuBi showing the electron ($\alpha$) and hole pockets ($\beta$ and $\gamma$).}
\label{bulk_LuBi}
\end{figure}

We calculate the bulk band structure of LuBi using LDA functional (Fig.~\ref{bulk_LuBi}(a)) along the high symmetry directions $W-L-\Gamma-X-W-K$ as shown in Fig.~\ref{struct}(b). Similar to LaBi, here also three doubly degenerate bands ($\alpha$, $\beta$, $\gamma$) pass the Fermi level $E_F$, giving rise to one electron pocket near the $X$-point and two hole pockets ($\beta$ and $\gamma$) at the $\Gamma$-point. However, the size of the electron and hole pockets are larger compared to LaBi as seen from the 3D Fermi surface plot of Fig.~\ref{bulk_LuBi}(c).

The conduction band and valence band are found to switch their orbital characters along the $\Gamma-X$ direction resulting in a band inversion. The $d_{xy}$ orbitals of Lu and $p$ orbitals of Bi participate in the band inversion and at each $k$-point there is a gap between the conduction and valence bands (gap $\sim$ 45 meV at the point of inversion).

However, the inclusion of mBJ exchange correlation shows a dramatic change in the band structure specially at the region of band inversion. It not only obliterates
the band inversion near the $X$ point, but also widens the gap ($\sim$ 250 meV) between the valence and the conduction band at the
point of inversion. The disappearance of band inversion along the $\Gamma-X$ direction indicates that, LuBi only has
an avoided band crossing in the bulk and thus is a conventional compensated semi-metal with extremely large magneto-resistance as discussed in section 3.3.

Calculation of $\mathbb{Z}_2$ index $\nu_0$ using Eq.~\ref{Z2} shows that, the $\mathbb{Z}_2$ topological invariant is zero for LuBi (as shown in Table \ref{tab2}) indicating its topologically trivial nature, in contrast to LaBi which is a strong topological semi-metal.
\subsection{Surface band structure }
\begin{figure}[h]
\centering
 $\begin{array}{cc}
\includegraphics[scale=0.087]{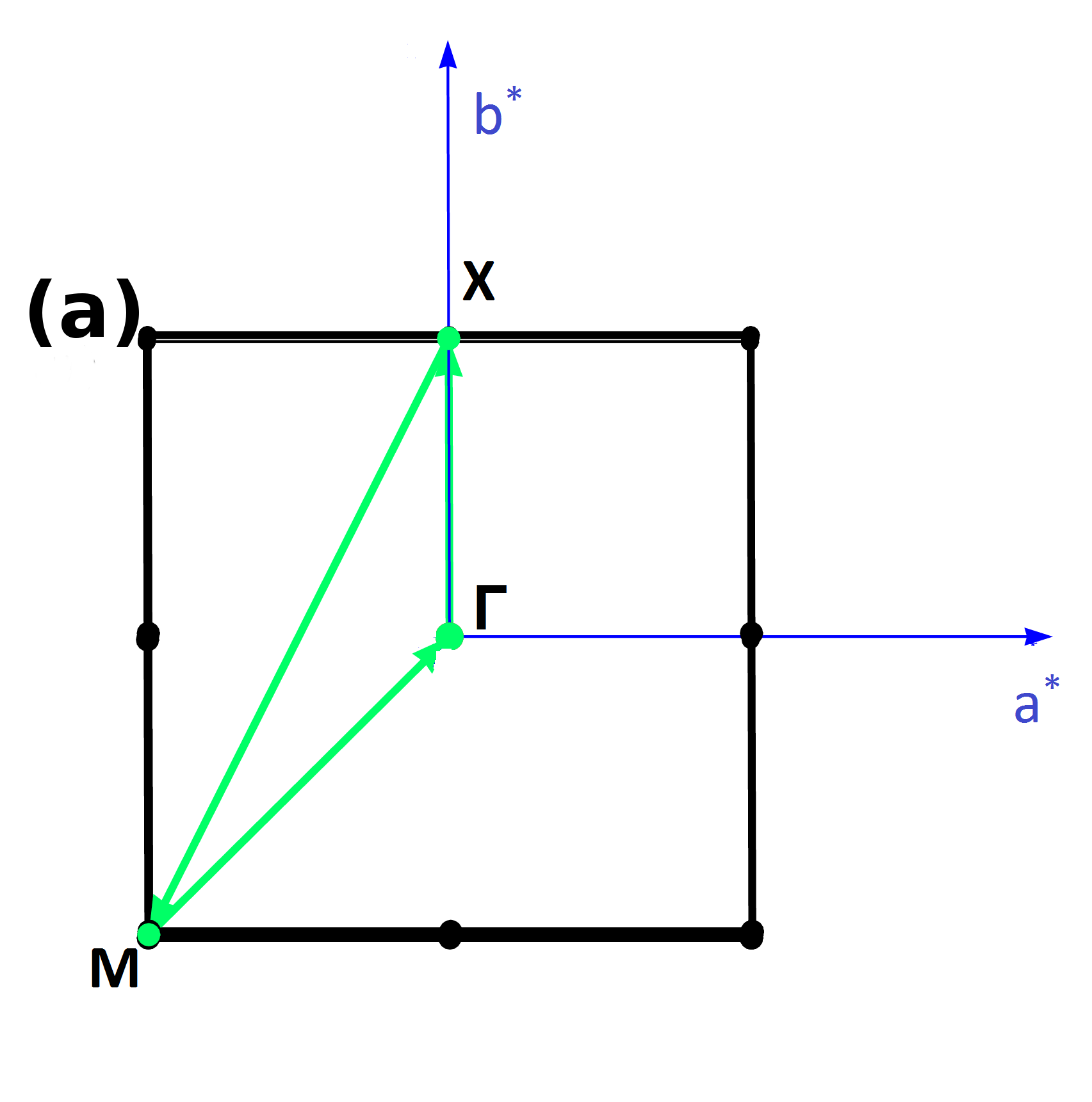}   &
\includegraphics[scale=.35]{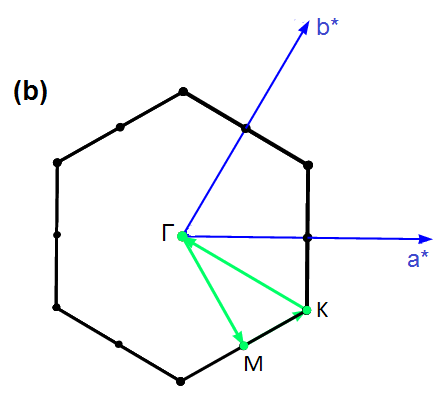} \\
\end{array}$
\caption{(a) The (001) projected 2D Brillouin zone of a fcc lattice with high symmetry points. (b) Projection of the (111) surface BZ. }
\label{SBZ}
\end{figure}

To examine the topological character of these materials, we calculate the (001) projected surface band
structures with a 16-layered slab in the (001) direction using supercell method. The (001) projected surface Brillouin zone (SBZ) is shown in Fig.~\ref{SBZ}(a).

\begin{figure}[h]
\centering
$\begin{array}{cccc}
\includegraphics[scale=.25]{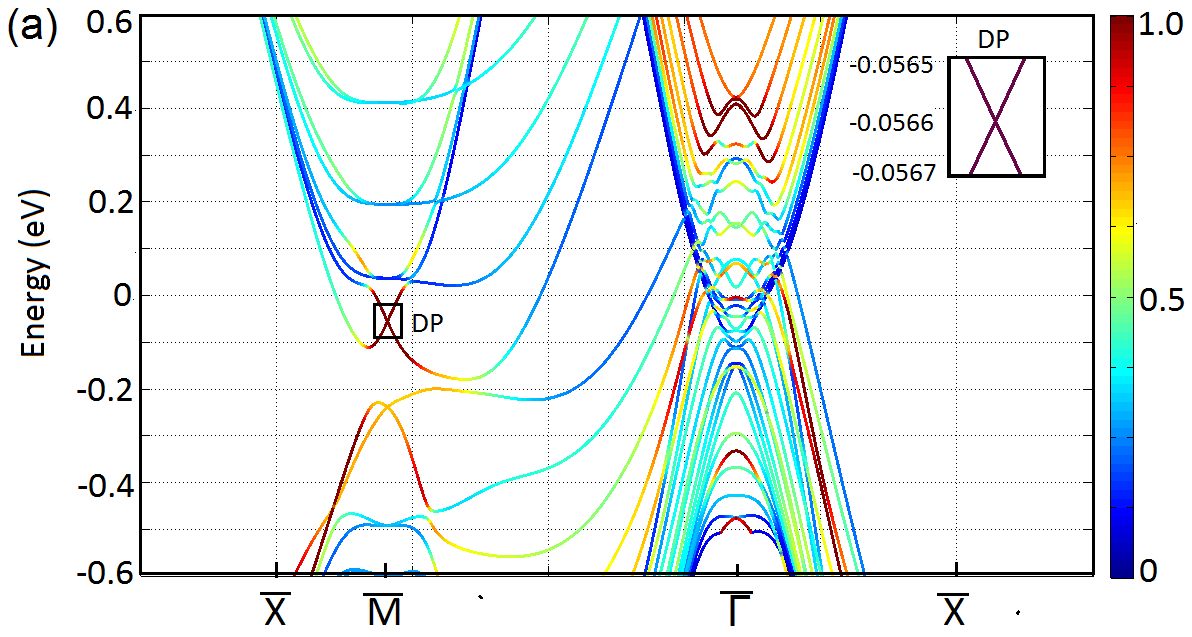}  &
\includegraphics[scale=.25]{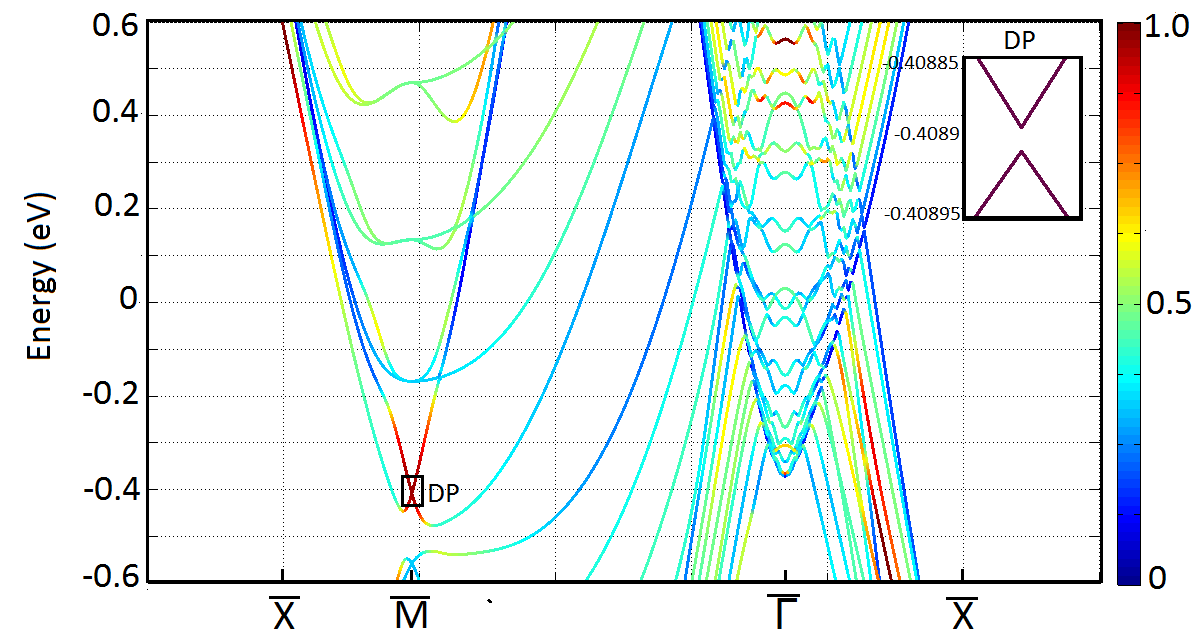} \\
\includegraphics[scale=.25]{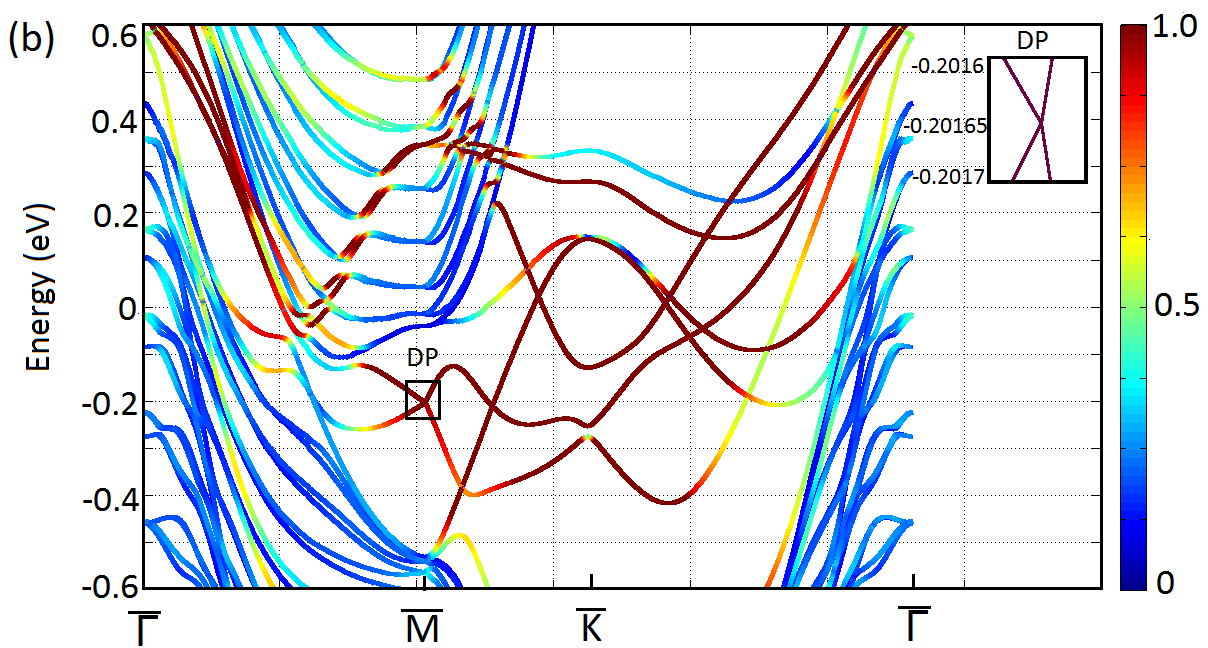} &
\includegraphics[scale=.25]{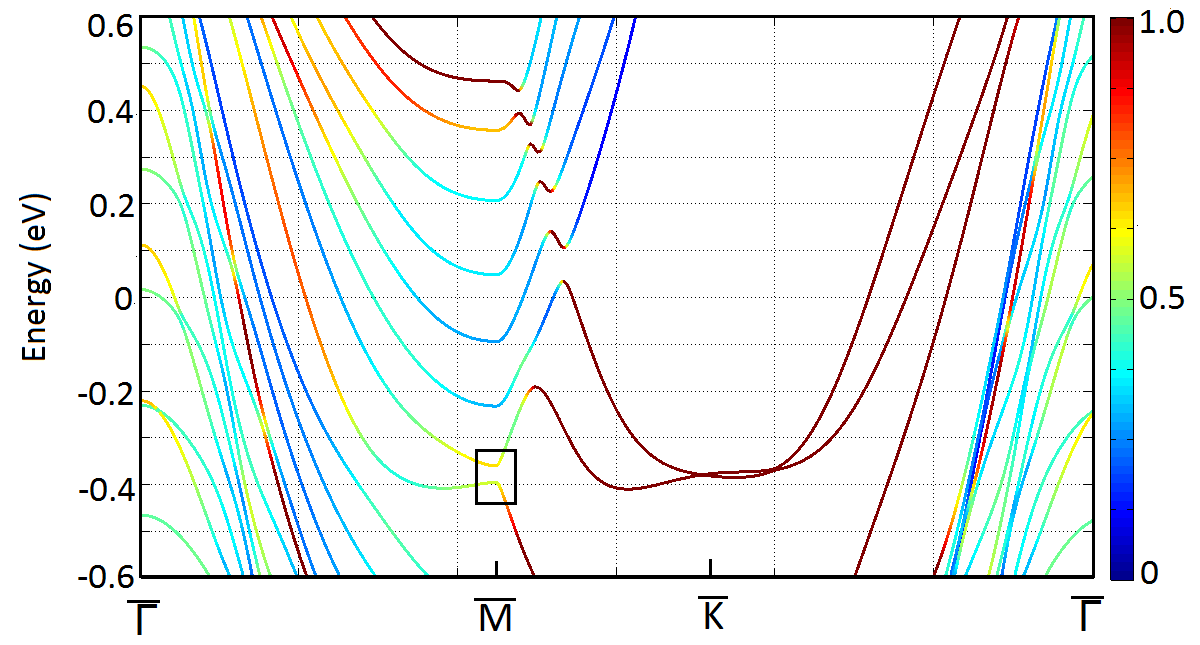} \\
\end{array}$
\caption{Surface band structures of LaBi and LuBi. Surface character of the bands are given by line intensity. The red curves represent the surface states whereas the blue curves give contribution of the projected bulk bands. (a) (001) surface band structure calculated using LDA for a slab of 16 atomic layers along the high symmetry directions as shown in Fig.~\ref{SBZ}(a). In LaBi (left), one gapless Dirac point (DP) exists at $\bar{M}$ and one massive Dirac cone appears at the $ \bar{\Gamma} $ point. However, in LuBi (right) the DP at the $\bar{M}$-point is gapped as shown in the inset. (b) The band structure of the (111) surface for a 36 layer slab. A single Dirac point appears near $\bar{M}$ in LaBi. However, in case of LuBi, a gap appears between the bands at $\bar{M}$. LaBi (left) and LuBi (right). Appearance of a single gapless DP on the (001) and (111) surfaces proves that LaBi is topologically non-trivial while for LuBi, it is gapped and indicates that LuBi is topologically trivial.}
\label{surface_band}
\end{figure}

The 16 atomic layers of the slab contains 32 atoms with 10 \r{A} vacuum separating the supercells in the real space.
Fig.~\ref{surface_band}(a) shows the (001) surface band structures where the bands are plotted along the
$ \left( \bar{X} - \bar{M} - \bar{\Gamma} - \bar{X} \right) $ direction. $ \Gamma $, $X$ and $L$ points of the bulk Brillouin zone are mapped
to $ \bar{\Gamma} $, $\bar{M} $ and $ \bar{X}$ points of the surface Brillouin zone. 
Our calculations using LDA functional show that one massless Dirac cone appears at the $\bar{M}$-point along with a gapped Dirac cone at $ \bar{\Gamma} $ in LaBi. The Dirac cone at the $\bar{M}$-point is shown in the inset of Fig.~\ref{surface_band}(a) (left). Dirac fermions acquire mass at $ \bar{\Gamma} $. Mass acquisition of Dirac fermions at the $ \bar{\Gamma}$ point in the (001) surface band dispersion was observed in LaBi by Y. Wu $\textit{et al}$~\cite{LuBi20} and R. Lou $\textit{et al}$~\cite{LuBi21}. Presence of single Dirac-like surface state indicates that LaBi is analogous to a $\mathbb{Z}_{2}$ topological semi-metal. 

On the other hand, in LuBi, a Dirac-cone-like surface state appears at the $\bar{M}$-point, as shown in Fig.~\ref{surface_band}(a) (right). However, close inspection of the DP shows that it is not closed i.e. a very small gap ( $\sim$ 0.01 meV) appears between the two bands.

Moreover, the calculated (111) surface dispersion of LaBi (Fig.~\ref{surface_band}(b) left) using a 36 layer slab shows the presence of a single Dirac cone at the $\bar{M}$ point~\cite{LuBi19}, which is absent in case of LuBi (Fig.~\ref{surface_band}(b) right). Since in LuBi, the surface DP is gapped in both (001) and (111) directions, it indicates that LuBi is topologically trivial and does not possess any topological character as inferred from the absence of band inversion in the mBJLDA calculation of the bulk band structure of LuBi.
\subsection{Charge compensation}
Using the LDA+SO calculation, we integrate the wannier-projected density of states of both LaBi and LuBi within the energy window of interest to find the carrier concentrations. Calculated carrier densities are shown in Table \ref{tab3}. The derived ratios of the no. of electrons and holes in these materials show almost perfect
compensation between electrons and hole charge carriers.\\
\indent In case of nonmagnetic materials with both types of carriers, the electrical resistivity in presence of a magnetic field B can be
derived from the semiclassical two-band model~\cite{LuBi38,LuBi39}:
\begin{equation}
\rho(B) = \frac{(n_{e}\mu_{e} + n_{h}\mu_{h}) + (n_{e}\mu_{h} + n_{h}\mu_{e})\mu_{e}\mu_{h}B^{2}}{e(n_{e}\mu_{e} + n_{h}\mu_{h})^{2} + e(\mu_{e}\mu_{h})^{2}(n_{e} - n_{h})^{2}B^{2}}
\end{equation}
where, $ \mu_{e} $ and $ \mu_{h} $ are the mobilities of electrons and holes respectively, e is the elementary electric charge.

\begin{table}[h!]
  \centering
  \caption{Concentration of electrons ($n_{e}$) and holes ($n_{h}$) in units of $ 10^{20} cm^{-3}$ calculated using LDA functional}
  \label{tab3}
  \begin{tabular}{c c c c c c}
    \hline
     & $n_{e}$ & $n_{h}$ ($\beta$) &  $n_{h}$ ($\gamma$)  & $ n_{h} = n_{h}$ ($\beta$) + $n_{h}$($\gamma$) & $ \frac{n_{e}}{n_{h}} $\\
    \hline
    LaBi & 2.48 & 0.44 & 1.61 & 2.05 & 1.21\\
    LuBi & 4.92 & 0.91 & 3.52 & 4.43 & 1.11 \\
    \hline

  \end{tabular}
\end{table}

Then, the change in resistance due to the external magnetic field or the magneto-resistance (MR) is given by~\cite{LuBi25},

\begin{equation}
\begin{split}
MR &= \frac{\rho(B) - \rho(0)}{\rho(0)} \\
&= \frac{n_{e}\mu_{e}n_{h}\mu_{h}(\mu_{e} + \mu_{h})^{2}B^{2}}{(n_{e}\mu_{e} + n_{h}\mu_{h})^{2} + (n_{e} - n_{h})^{2}(\mu_{e}\mu_{h})^{2}B^{2}}
\end{split}
\end{equation}

Now for a compensated material, $ n_{e} = n_{h} $, which yields
\begin{equation}
MR = \mu_{e}\mu_{h}B^{2}
\end{equation}

Therefore, the MR is proportional to the product of the mobilities of the carriers and it shows a quadratic dependence on the
magnetic field B. Again mobility of a charge carrier with effective mass $ m^{*} $ is proportional to $ \frac{\tau}{m^{*}} $,
where $\tau$ is the mean free time of the carrier. Our calculations (Table \ref{tab3}) show that the carrier densities are quite low in these materials.
Also, the bands, near $E_F$ are dispersive, which indicates the smaller effective mass of the carriers. As a result,
the carrier mobilities at low temperature would be high. Since magneto-resistance depends on the mobilities of electrons and holes, it is expected
that they should exhibit extremely large magneto-resistance (XMR), similar to the other compensated semi-metals like LuSb and $ WTe_{2}$~\cite{LuBi13,LuBi14,LuBi22,LuBi25,LuBi40}.

Again, the calculated carrier concentrations of LuBi are almost double than that of LaBi. It suggests that the
electrical conductivity of LuBi, which is proportional to the no. density of carriers, may be higher compared to LaBi.

\section{Conclusions}

In conclusion, using the $ \textit{ab initio} $ calculations, we have investigated the electronic structures of LaBi and LuBi with two different types of exchange-correlations (LDA and LDA+mBJ). While in case of LaBi, the band anticrossing is preserved along the $\Gamma-X$ direction with both the exchange-correlations, the inclusion of mBJ in LuBi results in an avoided band inversion. Thus the topological properties of LaBi are robust while we conclude from our analysis that LuBi is a conventional compensated semi-metal with the possibility of extremely high magneto-resistance. Hence the incorporation of mBJ over the conventional LDA (GGA) functionals, which leads to accurate band gap evaluation in case of semiconductors (insulators)~\cite{LuBi37a}, has some very interesting consequences in semi-metallic systems of topological importance. Presence of energy gap between conduction and valence bands at each $k$-point allows us to determine the $\mathbb{Z}_{2}$ invariant for LaBi and LuBi. Though we have not calculated electron-phonon coupling strengths in these materials, in general for topological insulators e.g. Bi$_2$Se$_3$, Bi$_2$Te$_3$, the strength of electron-phonon coupling is very weak~\cite{LuBi46,LuBi47} and their bulk band structures and topological surface states remain unaffected by lattice vibrations. So, we proceed to calculate the $\mathbb{Z}_{2}$ invariant for LaBi and LuBi using the formula proposed by Fu and Kane~\cite{LuBi45}. It is found that the $\mathbb{Z}_{2}$ index is 1 for LaBi, thus proving its non-trivial topological character. On the other hand, LuBi results in a zero $\mathbb{Z}_{2}$ index and is thus a conventional semi-metal.

Since the scope of the present investigation has topology within its ambit, the probe of the surface states is rendered indispensable. To that end, we have also calculated the surface band dispersions using the LDA functional. Although the surface states on the (001) and (111) surfaces show the presence of single Dirac cones in LaBi, gapped Dirac cones are observed at the $\bar{M}$ point in both (001) and (111) surface band structures of LuBi. Absence of massless Dirac cones on the (001) and (111) surfaces indicate that LuBi is a conventional semi-metal whereas in LaBi, gapless Dirac cones appear on both (001) and (111) surfaces, indicating the topologically non-trivial nature of LaBi. Though the Dirac points on the (001) and (111) surfaces of LaBi are below the Fermi level, we can tune the Fermi level to bring it to the position of the Dirac point by external (gating) or internal (doping) means in order to see the experimentally observable effects. The topological aspects of Bi-based materials owe their origin to the strong spin-orbit coupling. The carriers in the topological surface states (TSS) of these materials have their spins locked to their momentum. As a result, these states are insensitive to backscattering between states of opposite momentum and opposite spin and are robust against small perturbations as long as they are invariant under time reversal symmetry (TRS)~\cite{LuBi48}. These features render the TSS a robust conductivity. The TSS therefore hold immense promise for potential applications in spintronic devices, where long spin coherence is required~\cite{LuBi49}, and also for quantum computing applications~\cite{LuBi50}. 

Our calculated carrier densities of LaBi and LuBi reveals that electron-hole compensation is an intrinsic characteristics of both the materials. The calculated carrier concentrations are of the order $10^{-20}$/$cm^{3}$ which shows that LaBi and LuBi are compensated semi-metals. As a result, they should show extremely large magneto-resistance, as expected from the semiclassical two-band model. The present work not only probes the topological aspects of LuBi as compared to LaBi from different aspects, but also establishes the technologically quite attractive property of electron-hole compensation in LuBi in an unambiguous way.

\section*{Acknowledgements}
The author appreciates access to the computing facilities of the DST-FIST (phase-II) project installed in the Department of Physics, IIT Kharagpur, India. UD would like to thank Monodeep Chakraborty, Sumanta Tewari and A. Taraphder for useful discussions.

\section*{References}

\bibliography{LuBi}

\end{document}